\batchmode
\makeatletter
\def\input@path{{\string"D:/Research Document/Ongoing Research/DTDD_New_Doc/DTDD_Clustering/Draft_Writing/ArXiv_Version/\string"}}
\makeatother
\documentclass[twocolumn,english,conference]{IEEEtran}
\usepackage[T1]{fontenc}
\usepackage{babel}
\usepackage{textcomp}
\usepackage{amsmath}
\usepackage{amssymb}
\usepackage{graphicx}
\usepackage[pdftex,unicode=true,
 bookmarks=true,bookmarksnumbered=true,bookmarksopen=true,bookmarksopenlevel=1,
 breaklinks=false,pdfborder={0 0 0},pdfborderstyle={},backref=false,colorlinks=false]
 {hyperref}
\hypersetup{pdftitle={Your Title},
 pdfauthor={Your Name},
 pdfpagelayout=OneColumn, pdfnewwindow=true, pdfstartview=XYZ, plainpages=false}

\makeatletter

\ifCLASSOPTIONcompsoc
\usepackage[caption=false,font=normalsize,labelfont=sf,textfont=sf]{subfig}
\else
\usepackage[caption=false,font=footnotesize]{subfig}
\fi
\usepackage{cite}
\usepackage{amsthm}
\newtheorem{theorem}{\textbf{Theorem}}
\newtheorem{corollary}{\textbf{Corollary}}

\ifCLASSOPTIONcompsoc
\usepackage[caption=false,font=normalsize,labelfont=sf,textfont=sf]{subfig}
\else
\usepackage[caption=false,font=footnotesize]{subfig}
\fi

\ifCLASSOPTIONconference
\else
\fi

\IEEEoverridecommandlockouts
\ifCLASSOPTIONcompsoc
\usepackage[caption=false,font=normalsize,labelfont=sf,textfont=sf]{subfig}
\else
\usepackage[caption=false,font=footnotesize]{subfig}
\fi

\@ifundefined{showcaptionsetup}{}{%
 \PassOptionsToPackage{caption=false}{subfig}}
\usepackage{subfig}
\makeatother

\begin{document}
\title{Performance Analysis for Multi-Antenna Small Cell Networks with Clustered
Dynamic TDD}
\author{\IEEEauthorblockN{Hongguang Sun$^{\dagger,\mathsection}$, Howard
H. Yang$^{\ddagger}$, Xijun Wang$^{\star}$, Chao Xu$^{\dagger,\mathsection}$,
and Tony Q.S. Quek$^{\ddagger}$}\IEEEauthorblockA{$^{\dagger}$School
of Information Engineering, Northwest A\&F University, Yangling, Shaanxi,
China\\$^{\ddagger}$Information System Technology and Design Pillar,
Singapore University of Technology and Design, Singapore\\$^{\star}$School
of Electronics and Communication Engineering, Sun Yat-sen University,
Guangzhou, Guangdong, China\\$^{\mathsection}$Key Laboratory of
Agricultural Internet of Things, Ministry of Agriculture and Rural
Affairs, Yangling, Shaanxi, China}}
\maketitle
\begin{abstract}
Small cell networks with dynamic time-division duplex (D-TDD) have
emerged as a potential solution to address the asymmetric traffic
demands in 5G wireless networks. By allowing the dynamic adjustment
of cell-specific UL/DL configuration, D-TDD flexibly allocates percentage
of subframes to UL and DL transmissions to accommodate the traffic
within each cell. However, the unaligned transmissions bring in extra
interference which degrades the potential gain achieved by D-TDD.
In this work, we propose an analytical framework to study the performance
of multi-antenna small cell networks with clustered D-TDD, where cell
clustering is employed to mitigate the interference from opposite
transmission direction in neighboring cells. With tools from stochastic
geometry, we derive explicit expressions and tractable tight upper
bounds for success probability and network throughput. The proposed
analytical framework allows to quantify the effect of key system parameters,
such as UL/DL configuration, cluster size, antenna number, and SINR
threshold. Our results show the superiority of the clustered D-TDD
over the traditional D-TDD, and reveal the fact that there exists
an optimal cluster size for DL performance, while UL performance always
benefits from a larger cluster.
\end{abstract}

\section{{\normalsize{}Introduction}}

To satisfy the unprecedented high demands of data traffic, network
densification has been considered as one of the key technologies in
5G wireless networks \cite{NDTD,UDNT}. With the densely deployed
small cell access points (SAPs), not only the cell coverage, but also
the spatial reuse gain is significantly enhanced \cite{SCND}. To
further improve the network performance, the combination of SAPs and
multiple-input multiple-output (MIMO) technology is promising, where
the spatial multiplexing gain and/or diversity gain can be exploited
\cite{MTIW}. However, the dense deployment of SAPs increases the
variation of traffic demands among different cells, and the accommodation
to traffic fluctuation is essential. Dynamic time-division duplex
(D-TDD) has emerged as a potential solution to address the asymmetric
traffic demands \cite{DUDC}. Different from conventional static time-division
duplex (S-TDD), where all SAPs employ the same uplink/downlink (UL/DL)
configuration, D-TDD can efficiently support the asymmetric traffic
by allowing each SAP to dynamically adjust its UL/DL resources \cite{ODTD,TAAE}.
However, the flexibility is achieved at the expense of two new types
of inter-cell interference: the SAP-to-SAP interference and mobile
user to mobile user (MU-to-MU) interference.

To reduce the inter-cell interference, cell clustering has been proposed
as an effective approach \cite{CCBD,CBDD,APTS}. With the cell clustering
scheme, small cells in close proximity are classified into the same
cluster, and adopt S-TDD in a per-cluster basis. As such, SAPs within
the same cluster synchronize their transmissions, and the SAP-to-SAP
and MU-to-MU interference can be eliminated within the cluster. Although
the benefit of cell-clustering scheme has been evaluated in heterogeneous
network \cite{CCBD} or centralized radio access network (C-RAN) \cite{CBDD}
via simulations, an analytical framework is essential to fully understand
the performance of clustered D-TDD scheme, and capture the effect
of key network parameters. However, the spatial randomness of node's
geographical location and the resulting aggregated interference distribution
within a given cluster, put rigorous challenges for the development
of analytical framework \cite{HCNWS}.

With tools from stochastic geometry, prior works model node's spatial
irregularity by using classic spatial point process like Poisson Point
Process (PPP) in D-TDD networks, without considering the cell clustering
interference mitigation scheme \cite{PTAS,DEHC}. The authors in \cite{APTS}
first model the cell clustering scheme in a single-antenna small cell
D-TDD network, and use an approximation approach to compute the aggregated
interference from different clusters. With the extensive use of multi-antenna
technique, the multi-antenna small cell network with clustered D-TDD
is a promising architecture. In such a network, an analytical framework
is required to characterize the synergy of cell clustering and multi-antenna
techniques, so as to provide guidelines for the network design. However,
to the best of our knowledge, there is no previous literature evaluating
the D-TDD network by jointly considering the multi-antenna and cell
clustering techniques.

In this work, we develop an analytical framework to study the performance
of multi-antenna small cell networks operating clustered D-TDD. We
model the SAP and MU locations as two independent PPPs, and form each
cluster as a hexagon. Zero-forcing (ZF) beam-forming technique is
considered for both DL and UL transmissions to cancel the inter-user
interference within each cell. To reduce the computational complexity,
we use an approximate method to calculate the aggregated interference,
and derive tractable tight upper bounds for success probability and
network throughput. The proposed analytical framework allows to characterize
the effect of key network parameters, and provides guideline for the
optimal design of cell clustering scheme.

\section{{\normalsize{}System Model}}

\subsection{{\normalsize{}Network model}}

We consider a small cell network operating clustered D-TDD scheme,
where the spatial locations of SAPs and MUs follow two independent
homogeneous PPPs $\Phi_{\mathrm{s}}$ and $\Phi_{\mathrm{u}}$ with
intensities $\lambda_{\mathrm{s}}$ and $\lambda_{\mathrm{u}}$, respectively.
We adopt the nearest association policy where each MU connects to
its closest SAP. Let $n\in\{0,1,....,N\}$ be the number of MUs associated
with each SAP where we limit the maximum number of MUs served by each
SAP to $N$. According to the nearest association policy, the probability
density function (PDF) of the distance from a MU to its associated
SAP can be derived as $f_{r}(r)=2\pi\lambda_{\textrm{s}}r\exp(-\pi\lambda_{\textrm{s}}r^{2})$
\cite{ATAT}. Each SAP is equipped with $M$ antennas, while each
MU has a single antenna. The transmit power of SAP (to each MU) and
MU are defined as $P_{\textrm{s}}$ and $Q_{\textrm{u}}$, respectively.
The channel model consists of two attenuation components, namely large-scale
pathloss, and small-scale Rayleigh fading. Specifically, the pathloss
function is given by $g\left(\Vert x\Vert\right)=\frac{1}{\Vert x\Vert^{\alpha}}$
with $\alpha>2$ being the pathloss exponent, and the small-scale
Rayleigh fading with unit mean is given by $h\sim\textrm{exp(1)}$.

\subsection{{\normalsize{}Transmission Scheme}}

In this network, we consider a fully-loaded model where each SAP always
has data to transmit to all its associated MUs. We adopt the space
division multiple access (SDMA) scheme, where the maximum number of
MUs (associated to each SAP) $N$ does not surpass the number of antennas
$M$, i.e., $N\leq M$. As a result, all the MUs can be served by
its associated SAP simultaneously. To cancel the inter-user interference,
we adopt ZF pre-coding at the DL SAPs, and ZF receiver at the UL SAPs
with perfect knowledge of the channel state information (CSI). As
such, the channel power gain between an SAP and an MU is different
when the SAP acts as a serving SAP or interfering SAP in both DL and
UL mode, which is discussed in more details in Section III.

\begin{figure}[t]
\centering\includegraphics[scale=0.5]{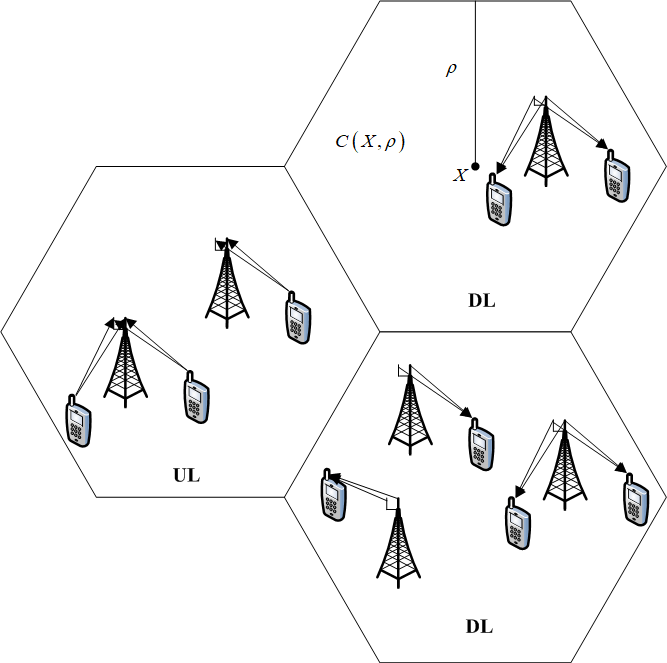}

\caption{An illu{\small{}stration of clustered D-TDD for two-antenna small
cell networks with }$\rho$ being the cluster radius{\small{}, where
SAPs located within the same cluster simultaneously align their UL/DL
configurations at each time slot.}}
\end{figure}

\subsection{Cell Clustering Scheme}

To reduce the interference from the opposite transmission direction,
the cell clustering scheme is adopted where we group Voronoi cells
that are closely located into clusters and align the transmissions
in each cluster, as depicted in Fig. 1.

To cover the whole network region without overlap, we form each cluster
as a hexagon. We define $C(T,\rho)$ as a hexagon centered at $T$
with $\rho$ being the cluster radius (from the hexagon center $T$
to the boundary). The area of $C(T,\rho)$ can be easily calculated
as $2\sqrt{3}\rho^{2}$, and the cluster size, i.e., the average number
of SAPs per cluster, is given by $l=2\sqrt{3}\rho^{2}\lambda_{s}$.
Note that the cluster size $l$ determines the operation pattern and
interference mechanism of the whole network. Obviously, as $l\rightarrow\infty$,
the network degenerates into operate under S-TDD scheme where all
SAPs have the same transmission direction, while $l\rightarrow1$
leads to the traditional D-TDD scheme where each SAP individually
sets its UL/DL configuration. With the proposed cell clustering scheme,
SAPs within the same cluster configure in DL (resp. UL) transmission
with probability $p_{\textrm{D}}$ (resp. 1 \textminus{} $p_{\textrm{D}}$).

\section{Performance Analysis}

\subsection{Success Probability}

We consider the constant bit-rate coding, and define $\gamma_{\textrm{D}}$
and $\gamma_{\textrm{U}}$ as the DL and UL SINR thresholds. With
the Slivyark\textquoteright s theorem \cite{SGWN,SGAI}, it is sufficient
to focus on the SINR of a typical MU or SAP that locates at the origin.
The success probability is defined as 
\begin{equation}
\mu_{\textrm{TX}}\triangleq\textrm{Pr}(\textrm{SINR}_{\textrm{TX}}>\gamma_{\textrm{TX}}),\:\textrm{TX}\in\{\textrm{D,U}\}.\label{eq:Suc_pro}
\end{equation}

With the cell clustering scheme, for a generic cluster, we can divide
the locations of SAPs $\Phi_{\mathrm{s}}$ (resp. MUs $\Phi_{\mathrm{u}}$)
into two independent PPPs $\Phi_{\mathrm{s}}^{\textrm{in}}$ and $\Phi_{\mathrm{s}}^{\textrm{out}}$
(resp. $\Phi_{\mathrm{u}}^{\textrm{in}}$ and $\Phi_{\mathrm{u}}^{\textrm{out}}$),
which lie within intra-cluster zone and cross-cluster zone, respectively.
By limiting the maximum number of MUs served by each SAP to $N$,
we can decompose $\Phi_{\mathrm{s}}^{\textrm{in}}$ and $\Phi_{\mathrm{s}}^{\textrm{out}}$
into $N$ tiers where the $n$-th ($n\in\{1,...,N\}$) tier is constituted
by SAPs with exactly $n$ associated MUs. According to this, we can
model the spatial locations of SAPs (MUs) in the $n$-th tier as $\Phi_{\mathrm{s}_{n}}^{\textrm{in}}$
($\Phi_{\mathrm{u}_{n}}^{\textrm{in}}$) and $\Phi_{\mathrm{s}_{n}}^{\textrm{out}}$
($\Phi_{\mathrm{u}_{n}}^{\textrm{out}}$), respectively. To analyze
the aggregated interference from cross-cluster zone, we approximate
the cross-cluster zone into a sequence of disjoint hexagonal rings.
Specifically, let $C(T,\rho)$ be the cluster of the typical SAP or
MU, the cross-cluster zone can be approximated by $\{\mathcal{A}_{k}\}_{k=1}^{\infty}=\{C(T,\sqrt{k+1}\rho)\backslash C(T,\sqrt{k}\rho)\}_{k=1}^{\infty}$
\cite{AAFM} where the $k$-th hexagonal ring $\mathcal{A}_{k}$ corresponds
to the $k$-th cluster, and the area of $\mathcal{A}_{k}$ is the
same as $C(T,\rho)$, i.e., $2\sqrt{3}\rho^{2}$. In the $k$-th ring,
we define $\Phi_{\mathrm{s}_{k}}^{\textrm{out}_{k}}$ and $\Phi_{\mathrm{u}_{k}}^{\textrm{out}_{k}}$
as the spatial locations of SAPs and MUs, respectively, and we have

{\small{}
\begin{equation}
\Phi_{\mathrm{s}}=\Phi_{\mathrm{s}}^{\textrm{in}}\cup\Phi_{\mathrm{s}}^{\textrm{out}}=\left(\cup_{n=1}^{N}\Phi_{\mathrm{s}_{n}}^{\textrm{in}}\right)\cup\left(\cup_{k=1}^{\infty}\cup_{n=1}^{N}\Phi_{\mathrm{s}_{n}}^{\textrm{out}_{k}}\right),\label{eq:Phi_s}
\end{equation}
\begin{equation}
\Phi_{\mathrm{u}}=\Phi_{\mathrm{u}}^{\textrm{in}}\cup\Phi_{\mathrm{u}}^{\textrm{out}}=\left(\cup_{n=1}^{N}\Phi_{\mathrm{u}_{n}}^{\textrm{in}}\right)\cup\left(\cup_{k=1}^{\infty}\cup_{n=1}^{N}\Phi_{\mathrm{u}_{n}}^{\textrm{out}_{k}}\right).\label{eq:Phi_u}
\end{equation}
}{\small\par}

For the typical MU located at the origin $\mathbf{0}$ in DL mode,
the received SINR can be expressed as
\begin{equation}
\textrm{\ensuremath{\textrm{SINR}_{\textrm{D}}}}=\frac{P_{\mathrm{s}}r_{0}^{-\alpha}\parallel\mathbf{h}_{0,0}^{\dagger}\mathbf{w}_{0,0}\parallel^{2}}{I_{\textrm{D}}+\sigma^{2}},\label{eq:DL_SINR}
\end{equation}
where $\mathbf{h}_{0,0}\in\mathbb{C}^{M\times1}$ and $\mathbf{w}_{0,0}\in\mathbb{C}^{M\times n_{0}}$,
respectively, denote the channel vector and the ZF pre-coding matrix
at the tagged DL SAP (to the typical MU), and $I_{\textrm{D}}$ denotes
the aggregated interference from DL SAPs and UL MUs, given by{\small{}
\begin{eqnarray}
I_{\textrm{D}} & = & \sum_{n=1}^{N}\sum_{x_{i}\in\Phi_{\mathrm{s}_{n}}^{\textrm{in}}\backslash\{x_{0}\}}P_{\mathrm{s}}\parallel x_{0,i}\parallel^{-\alpha}\mid\mathbf{h}_{0,i}^{\dagger}\mathbf{W}_{i}\mid^{2}\nonumber \\
 & + & \sum_{k=1}^{\infty}\biggl[\boldsymbol{1}_{\{\textrm{TX=D}\}}^{\textrm{out}_{k}}\sum_{n=1}^{N}\sum_{x_{j}\in\Phi_{\mathrm{s}_{n}}^{\textrm{out}_{k}}}P_{\mathrm{s}}\parallel x_{0,j}\parallel^{-\alpha}\mid\mathbf{h}_{0,j}^{\dagger}\mathbf{W}_{j}\mid^{2}\nonumber \\
 & + & \boldsymbol{1}_{\{\textrm{TX=U}\}}^{\textrm{out}_{k}}\sum_{n=1}^{N}\sum_{z_{l}\in\Phi_{\mathrm{u}_{n}}^{\textrm{out}_{k}}}Q_{\mathrm{u}}\parallel z_{0,l}\parallel^{-\alpha}\mid h_{0,l}\mid^{2}\biggr],\label{eq:I_D}
\end{eqnarray}
}where indicator functions $\boldsymbol{1}_{\{\textrm{TX=D}\}}^{\textrm{out}_{k}}$
and $\boldsymbol{1}_{\{\textrm{TX=U}\}}^{\textrm{out}_{k}}$ represent
that the SAPs in the $n$-th cluster are configured in DL and UL direction,
respectively. The first part, second part and last part in \eqref{eq:I_D}
denote the DL interference from intra-cluster zone, cross-cluster
zone, and UL interference from cross-cluster zone, respectively. The
channel vectors from the interfering DL SAP located at $x_{i}$ and
$x_{j}$ are denoted by $\mathbf{h}_{0,i}\in\mathbb{C}^{M\times1}$
and $\mathbf{h}_{0,j}\in\mathbb{C}^{M\times1}$, respectively. For
the UL interfering MUs, the channel gain from $z_{l}$ to the typical
MU is denoted by $h_{0,l}\sim\mathcal{CN}(0,1)$. We define $\tilde{\mathbf{H}}_{i}=[\mathbf{\tilde{h}}_{0,i},...,\mathbf{\tilde{h}}_{c,i},...,\mathbf{\tilde{h}}_{n-1,i}]^{\dagger}\in\mathbb{C}^{n\times M}$
as the channel matrix between a DL interfering SAP and all the $n$
associated MUs. The direction of each vector channel is represented
by $\mathbf{\tilde{h}}_{c,i}\triangleq\frac{\mathbf{h}_{c,i}}{\parallel\mathbf{h}_{c,i}\parallel}$,
where $\parallel\cdot\parallel$ denotes the Euclidean norm. By using
ZF pre-coding, the columns of the pre-coding matrix $\mathbf{W}_{i}=[\mathbf{w}_{i,c}]_{1\leq c\leq n}\in\mathbb{C}^{M\times n}$
are exactly the columns of the pseudo-inverse $\mathbf{H}_{i}^{\mathtt{\mathsf{H}}}\triangleq\mathbf{\tilde{H}_{\mathit{i}}^{\dagger}}(\tilde{\mathbf{H}_{i}}\mathbf{\tilde{H}_{\mathit{i}}^{\dagger}})^{-1}\in\mathbb{C}^{M\times n}$,
where $(.)^{\mathsf{H}}$ and $(.)^{\dagger}$ represent the pseudo-inverse
and conjugate transpose, respectively.

Let $n_{0}\in\{1,2,...,N\}$ be the number of MUs associated with
the tagged DL SAP. The desired channel power gain can be derived as
$h_{0,0}^{\textrm{D}}=\parallel\mathbf{h}_{0,0}^{\dagger}\mathbf{w}_{0,0}\parallel^{2}\sim\Gamma(M-n_{0}+1,1)$
\cite{DMHM}, the DL interference channel power gain is given by $g_{\mathbf{x}_{i},\mathrm{SAP}}^{\textrm{D}}=\mid\mathbf{h}_{0,i}^{\dagger}\mathbf{W}_{i}\mid^{2}\sim\Gamma(n,1)$,
and the UL interference channel power gain is derived as $g_{\mathrm{MU}}^{\textrm{D}}=\mid h_{0,l}\mid^{2}\sim\textrm{Exp}(1)$.

For the typical SAP located at the origin $\mathbf{0}$ in UL mode,
the received SINR can be expressed as
\begin{equation}
\textrm{\ensuremath{\textrm{SINR}_{\textrm{U}}}}=\frac{Q_{\mathrm{u}}r_{0}^{-\alpha}\parallel\mathbf{v}_{0}^{\dagger}\mathbf{g}_{0,0}\parallel^{2}}{I_{\textrm{U}}+\mid\mathbf{v}_{0}^{\dagger}\mathbf{\boldsymbol{n}_{0}}\mid^{2}},
\end{equation}
where $\mathbf{g}_{0,0}\in\mathbb{C}^{M\times1}$ and $\mathbf{v}_{0}\in\mathbb{C}^{M\times1}$,
respectively, denote the channel vector from the expected UL MU located
at $z_{0}$ to the typical SAP, and the unit norm ZF receive filter.
The noise power {\small{}$\mid\mathbf{v}_{0}^{\dagger}\mathbf{\boldsymbol{n}_{0}}\mid^{2}=\sigma^{2},$}
and the aggregated interference $I_{\textrm{U}}$ is given by{\small{}
\begin{eqnarray}
I_{\textrm{U}} & = & \sum_{n=1}^{N}\sum_{z_{i}\in\Phi_{\mathrm{u}_{n}}^{\textrm{in}}\backslash\{z_{0}\}}Q_{\mathrm{u}}\parallel z_{0,i}\parallel^{-\alpha}\mid\mathbf{v}_{0}^{\dagger}\mathbf{g}_{0,i}\mid^{2}\nonumber \\
 & + & \sum_{k=1}^{\infty}\biggl[\boldsymbol{1}_{\{\textrm{TX=U}\}}^{\textrm{out}_{k}}\sum_{n=1}^{N}\sum_{z_{m}\in\Phi_{\mathrm{u}_{n}}^{\textrm{out}_{k}}}Q_{\mathrm{u}}\parallel z_{0,m}\parallel^{-\alpha}\mid\mathbf{v}_{0}^{\dagger}\mathbf{g}_{0,m}\mid^{2}\nonumber \\
 & + & \boldsymbol{1}_{\{\textrm{TX=D}\}}^{\textrm{out}_{k}}\sum_{n=1}^{N}\sum_{x_{j}\in\Phi_{\mathrm{s}_{n}}^{\textrm{out}_{k}}}P_{s}\parallel x_{0,j}\parallel^{-\alpha}\mid\mathbf{v}_{0}^{\dagger}\mathbf{\boldsymbol{\mathcal{H}}}_{0,j}^{\dagger}\mathbf{W}_{j}\mid^{2}\biggr],\label{eq:I_U}
\end{eqnarray}
}where the first part, second part and last part denote the UL interference
from intra-cluster zone, cross-cluster zone, and DL interference from
cross-cluster zone, respectively. The symbols $\mathbf{g}_{0,i}\in\mathbb{C}^{M\times1}$
and $\mathbf{\boldsymbol{\mathcal{H}}}_{0,j}^{\dagger}\in\mathbb{C}^{M\times M}$
denote the channel vectors from the interfering MU located at $z_{i}$
and SAP located at $x_{j}$ (to the typical SAP), respectively.

Let $n_{0}\in\{1,2,...,N\}$ be the number of MUs associated with
the typical UL SAP. By using ZF receiver, a unit norm receive filter
$\mathbf{v}_{0}\in\mathbb{C}^{M\times1}$ is selected orthogonal to
the channel vectors of other $n_{0}-1$ interferer associated with
the same typical SAP, i.e., $\mid\mathbf{v}_{0}^{\dagger}\mathbf{g}_{0,p}\mid^{2}=0$
for $p=1,...,n_{0}-1$.\footnote{It is worth noting that when $n_{0}=1$, the ZF receiver becomes the
maximal ratio combining (MRC) technique.} As is derived in \cite{MACA}, we have $h_{0,0}^{\textrm{U}}=\parallel\mathbf{v}_{0}^{\dagger}\mathbf{g}_{0,0}\parallel^{2}\sim\Gamma(M-n_{0}+1,1)$,
$g_{\mathrm{MU}}^{\textrm{U,in}}=\mid\mathbf{v}_{0}^{\dagger}\mathbf{g}_{0,i}\mid^{2}\sim\textrm{Exp}(1)$,
$g_{\mathrm{MU}}^{\textrm{U,out}}=\mid\mathbf{v}_{0}^{\dagger}\mathbf{g}_{0,m}\mid^{2}\sim\textrm{Exp}(1)$,
and $g_{\mathbf{x}_{j},\mathrm{SAP}}^{\textrm{U}}=\mid\mathbf{v}_{0}^{\dagger}\mathbf{\boldsymbol{\mathcal{H}}}_{0,j}^{\dagger}\mathbf{W}_{j}\mid^{2}\thicksim\Gamma(n,1)$,
respectively.

\begin{theorem}

The DL and UL success probabilities of the typical SAP with $n_{0}\in\{1,2,...,N\}$
associated MUs can be approximated by{\small{}
\begin{align}
\mu_{\textrm{D}}(n_{0}) & \approx\int_{0}^{\rho}2\pi\lambda_{s}r_{0}\exp(-\pi\lambda_{s}r_{0}^{2})\biggl[\sum_{i=0}^{M-n_{0}}\frac{1}{i!}(-s)^{i}\nonumber \\
 & \qquad\qquad\qquad\qquad\qquad\times\frac{d^{i}}{ds^{i}}\mathcal{L}_{I_{IN}^{\textrm{D}}}(s)\biggr]_{s=\frac{\gamma_{\mathtt{\textrm{D}}}r_{0}^{\alpha}}{P_{\mathrm{s}}}}\mathrm{d}r_{0},\label{eq:mu_D}
\end{align}
\begin{align}
\mu_{\textrm{U}}(n_{0}) & \approx\int_{0}^{\rho}2\pi\lambda_{s}r_{0}\exp(-\pi\lambda_{s}r_{0}^{2})\biggl[\sum_{i=0}^{M-n_{0}}\frac{1}{i!}(-s)^{i}\nonumber \\
 & \qquad\qquad\qquad\qquad\qquad\times\frac{d^{i}}{ds^{i}}\mathcal{L}_{I_{IN}^{\textrm{U}}}(s)\biggr]_{s=\frac{\gamma_{\mathtt{\textrm{U}}}r_{0}^{\alpha}}{Q_{\mathrm{u}}}}\mathrm{d}r_{0},\label{eq:mu_U}
\end{align}
}where $\mathcal{L}_{I_{IN}^{\textrm{D}}}(s)$ and $\mathcal{L}_{I_{IN}^{\textrm{U}}}(s)$
are given by{\footnotesize{}
\begin{eqnarray}
 &  & \mathcal{L}_{I_{IN}^{\textrm{D}}}(s)\nonumber \\
 & = & \exp\bigl(-s\sigma^{2}\bigr)\exp\biggl\{-2\pi\sum_{n=1}^{N}\lambda_{s,n}\sum_{l=1}^{n}C_{n}^{l}\int_{r_{0}}^{\rho}\frac{\bigl(sP_{\mathrm{s}}r^{-\alpha}\bigr)^{l}r}{(1+sP_{\mathrm{s}}r^{-\alpha})^{n}}\mathrm{d}r\biggr\}\nonumber \\
 & \times & \prod_{k=1}^{\infty}\biggl\{ p_{\textrm{D}}\exp\biggl(-2\pi\sum_{n=1}^{N}\lambda_{\textrm{s},n}\sum_{l=1}^{n}C_{n}^{l}\int_{\sqrt{k}\rho}^{\sqrt{k+1}\rho}\frac{(sP_{\mathrm{s}}r^{-\alpha})^{l}r}{(1+sP_{\mathrm{s}}r^{-\alpha})^{n}}\mathrm{d}r\biggr)\nonumber \\
 &  & +\bigl(1-p_{\textrm{D}}\bigr)\exp\Bigl(-2\pi\sum_{n=1}^{N}\lambda_{\textrm{u},n}\biggl(\Theta\Bigl(\alpha,1,\bigl(sQ_{u}\bigr)^{-1},\sqrt{k+1}\rho\Bigr)\nonumber \\
 &  & \qquad\qquad\qquad\qquad\qquad\qquad-\Theta\Bigl(\alpha,1,\bigl(sQ_{u}\bigr)^{-1},\sqrt{k}\rho\Bigr)\biggr)\biggr)\biggr\},\label{eq:Laplace_DL}
\end{eqnarray}
\begin{eqnarray}
 &  & \mathcal{L}_{I_{IN}^{\textrm{U}}}(s)\nonumber \\
 & = & \exp\bigl(-s\sigma^{2}\bigr)\exp\Bigl(-2\pi\sum_{n=1}^{N}\lambda_{\textrm{u},n}\Theta\Bigl(\alpha,1,\bigl(sQ_{\textrm{u}}\bigr)^{-1},\rho\Bigr)\Bigr)\nonumber \\
 & \times & \prod_{k=1}^{\infty}\biggl\{ p_{\textrm{D}}\exp\biggl(-2\pi\sum_{n=1}^{N}\lambda_{\textrm{s},n}\sum_{l=1}^{n}C_{n}^{l}\int_{\sqrt{k}\rho}^{\sqrt{k+1}\rho}\frac{(sP_{\mathrm{s}}r^{-\alpha})^{l}r}{(1+sP_{\mathrm{s}}r^{-\alpha})^{n}}\mathrm{d}r\biggr)\nonumber \\
 &  & +\bigl(1-p_{\textrm{D}}\bigr)\exp\Bigl(-2\pi\sum_{n=1}^{N}\lambda_{\textrm{u},n}\biggl(\Theta\Bigl(\alpha,1,\bigl(sQ_{\textrm{u}}\bigr)^{-1},\sqrt{k+1}\rho\Bigr)\nonumber \\
 &  & \qquad\qquad\qquad\qquad\qquad\qquad-\Theta\Bigl(\alpha,1,\bigl(sQ_{\textrm{u}}\bigr)^{-1},\sqrt{k}\rho\Bigr)\biggr)\biggr\}\label{eq:Laplace_UL}
\end{eqnarray}
}with $\lambda_{\textrm{s},n}=\lambda_{s}f(n)$, $\lambda_{\textrm{u},n}=\lambda_{s}f(n)\cdot n$
and 
\begin{align*}
\Theta\Bigl(\alpha,\beta,u,d\Bigr) & \triangleq\int_{0}^{d}\frac{r^{\beta}}{1+ur^{\alpha}}dr\\
 & =\frac{d^{\beta+1}}{\beta+1}_{2}F_{1}(1,\frac{\beta+1}{\alpha};1+\frac{\beta+1}{\alpha};-ud^{\alpha})
\end{align*}
 with $_{2}F_{1}(\cdot,\cdot;\cdot;\cdot)$ being the Gaussian hyper-geometric
function.
\begin{IEEEproof}
See Appendix A.
\end{IEEEproof}
\end{theorem}

As can be seen from \eqref{eq:mu_D} and \eqref{eq:mu_U}, the computation
of $\mu_{\textrm{D}}(n_{0})$ and $\mu_{\textrm{U}}(n_{0})$ requires
evaluating higher order derivatives of the Laplace transform. A larger
parameter $M-n_{0}$ leads to higher computational complexity. In
the following corollary, we derive the upper bounds of $\mu_{\textrm{D}}(n_{0})$
and $\mu_{\textrm{U}}(n_{0})$ by employing the complementary cumulative
distribution function (CCDF) of Gamma distribution.

\begin{corollary}

The DL and UL success probabilities of the typical SAP with $n_{0}\in\{1,2,...,N\}$
associated MUs are upper bounded by{\small{}
\begin{align}
\mu_{\textrm{D}}(n_{0}) & \leq\int_{0}^{\rho}2\pi\lambda_{s}r_{0}\exp(-\pi\lambda_{s}r_{0}^{2})\biggl[\sum_{i=1}^{\Delta_{n_{0}}}(-1)^{i+1}C_{\Delta_{n_{0}}}^{i}\nonumber \\
 & \times\mathcal{L}_{I_{IN}^{\textrm{D}}}(i\cdot(\Gamma(\Delta_{n_{0}}+1))^{-\frac{1}{\Delta_{n_{0}}}}\cdot s)\biggr]_{s=\frac{\gamma_{\mathtt{\textrm{D}}}r_{0}^{\alpha}}{P_{\mathrm{s}}}}\mathrm{d}r_{0},\label{eq:mu_D-upper-1}
\end{align}
\begin{align}
\mu_{\textrm{U}}(n_{0}) & \leq\int_{0}^{\rho}2\pi\lambda_{s}r_{0}\exp(-\pi\lambda_{s}r_{0}^{2})\biggl[\sum_{i=1}^{\Delta_{n_{0}}}(-1)^{i+1}C_{\Delta_{n_{0}}}^{i}\nonumber \\
 & \times\mathcal{L}_{I_{IN}^{\textrm{U}}}(i\cdot(\Gamma(\Delta_{n_{0}}+1))^{-\frac{1}{\Delta_{n_{0}}}}\cdot s)\biggr]_{s=\frac{\gamma_{\mathtt{\textrm{U}}}r_{0}^{\alpha}}{Q_{\mathrm{u}}}}\mathrm{d}r_{0},\label{eq:mu_U-upper-1}
\end{align}
}where $\Delta_{n_{0}}\triangleq M-n_{0}+1$, $\mathcal{L}_{I_{IN}^{\textrm{D}}}(s)$
and $\mathcal{L}_{I_{IN}^{\textrm{U}}}(s)$ are given by \eqref{eq:Laplace_DL}
and \eqref{eq:Laplace_UL} in Theorem 1, respectively.
\begin{IEEEproof}
See Appendix B.
\end{IEEEproof}
\end{corollary}

With the law of total probability, the overall DL and UL success probabilities
can be derived as
\begin{equation}
\mu_{\textrm{D}}=\sum_{n_{0}=1}^{N}\mu_{\textrm{D}}(n_{0})f(n_{0}),
\end{equation}
\begin{equation}
\mu_{\textrm{U}}=\sum_{n_{0}=1}^{N}\mu_{\textrm{U}}(n_{0})f(n_{0}).
\end{equation}

\subsection{Network Throughput}

With the success probability obtained above, we derive the network
throughput of the small cell network in this subsection. Conditioned
on the DL fraction $p_{\textrm{D}}$ and the cluster size $l$, the
DL and UL network throughput (in $Bits/Sec/Hz/m^{2}$) can be written
as
\begin{equation}
\mathcal{T}_{\textrm{D}}=p_{\textrm{D}}\lambda_{s}\log_{2}(1+\gamma_{\textrm{D}})\sum_{n_{0}=1}^{N}\mu_{\textrm{D}}(n_{0})f(n_{0})n_{0},
\end{equation}
\begin{equation}
\mathcal{T}_{\textrm{U}}=(1-p_{\textrm{D}})\lambda_{s}\log_{2}(1+\gamma_{\textrm{U}})\sum_{n_{0}=1}^{N}\mu_{\textrm{U}}(n_{0})f(n_{0})n_{0}.
\end{equation}

\section{Numerical Results}

In this section, we first verify the theoretical model by means of
simulations, and then provide key design insights for multi-antenna
small cell networks under clustered D-TDD. We perform all simulations
over a square window of 1000 \texttimes{} 1000 $m^{2}$ with 10000
iterations. With the clustered D-TDD scheme, each SAP synchronizes
its transmission to all the other SAPs within the same cluster, and
serves all its associated MUs. Due to the high computational complexity
of the analytical results derived in \eqref{eq:mu_D} and \eqref{eq:mu_U},
in the following figures, we only plot the upper bound of the analytical
results by using \eqref{eq:mu_D-upper-1} and \eqref{eq:mu_U-upper-1},
respectively. Unless otherwise specified, we use the following default
values of network parameters: SAP density $\lambda_{s}=10^{-3}$$m^{-2}$,
MU density $\lambda_{u}=10\lambda_{s}$, transmit power of MU $Q_{\textrm{u}}=17$
dBm, maximum number of MUs associated to each SAP $N=3$, and the
number of antennas equipped by each SAP $M=8$.

\begin{figure}[t]
\centering\includegraphics[scale=0.5]{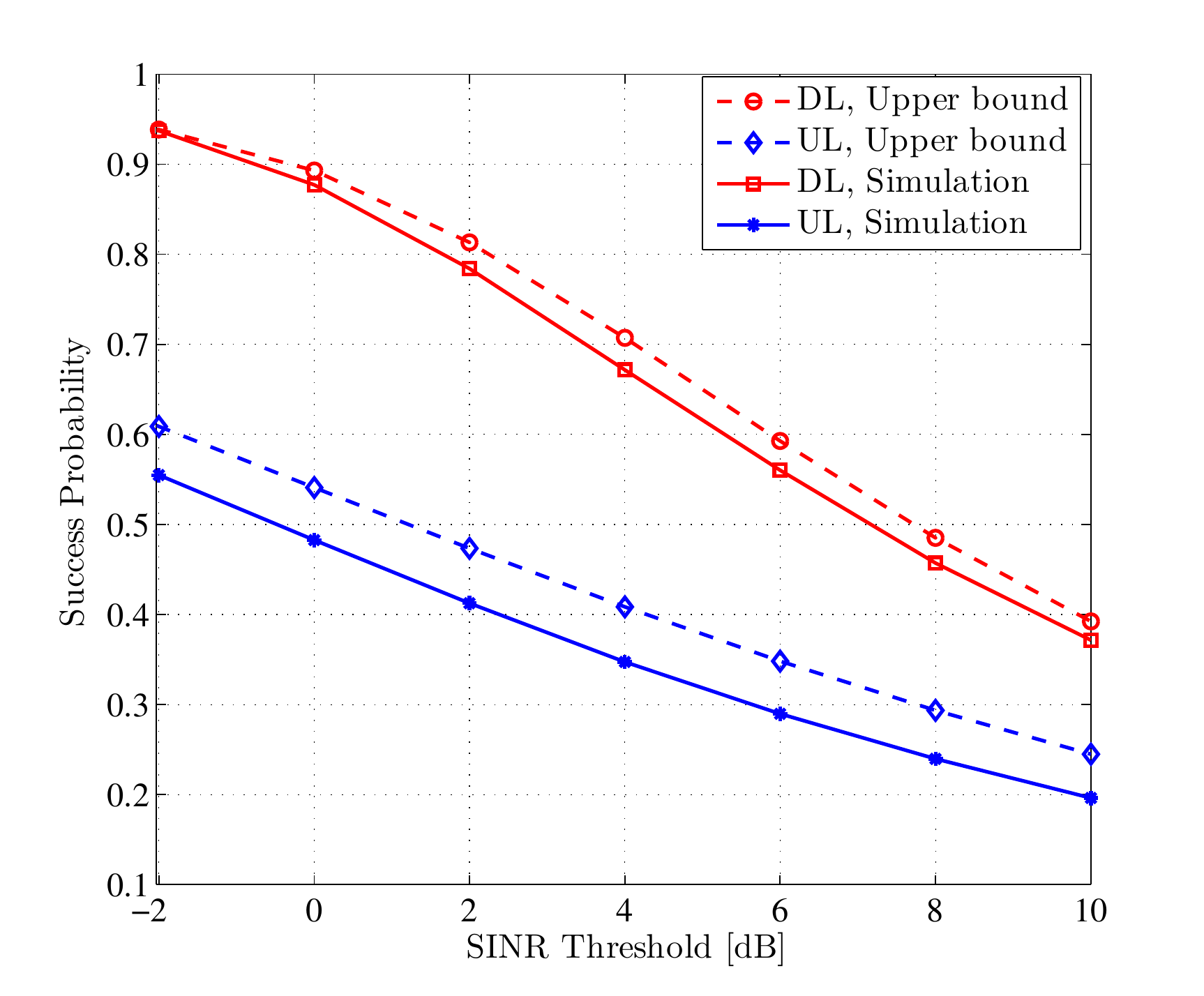}

\caption{Success probability as a function of SINR threshold with cluster size
$l=3$, antenna number per SAP $M=8$, and DL fraction $p_{\textrm{D}}=0.5$.}
\end{figure}

Figure 2 depicts the success probability as a function of SINR threshold.
We observe that the upper bound of DL success probability is much
tighter than that of UL success probability. This can be explained
by the use of PPP approximation for active MUs in UL mode. Note that
with the limitation on the maximum number of MUs $N$ (associated
with each SAP) and the DL fraction $p_{\textrm{D}}$, the active UL
interfering MUs do not distribute as a PPP.

\begin{figure}[t]
\centering%
\subfloat[]%
{\centering\includegraphics[scale=0.5]{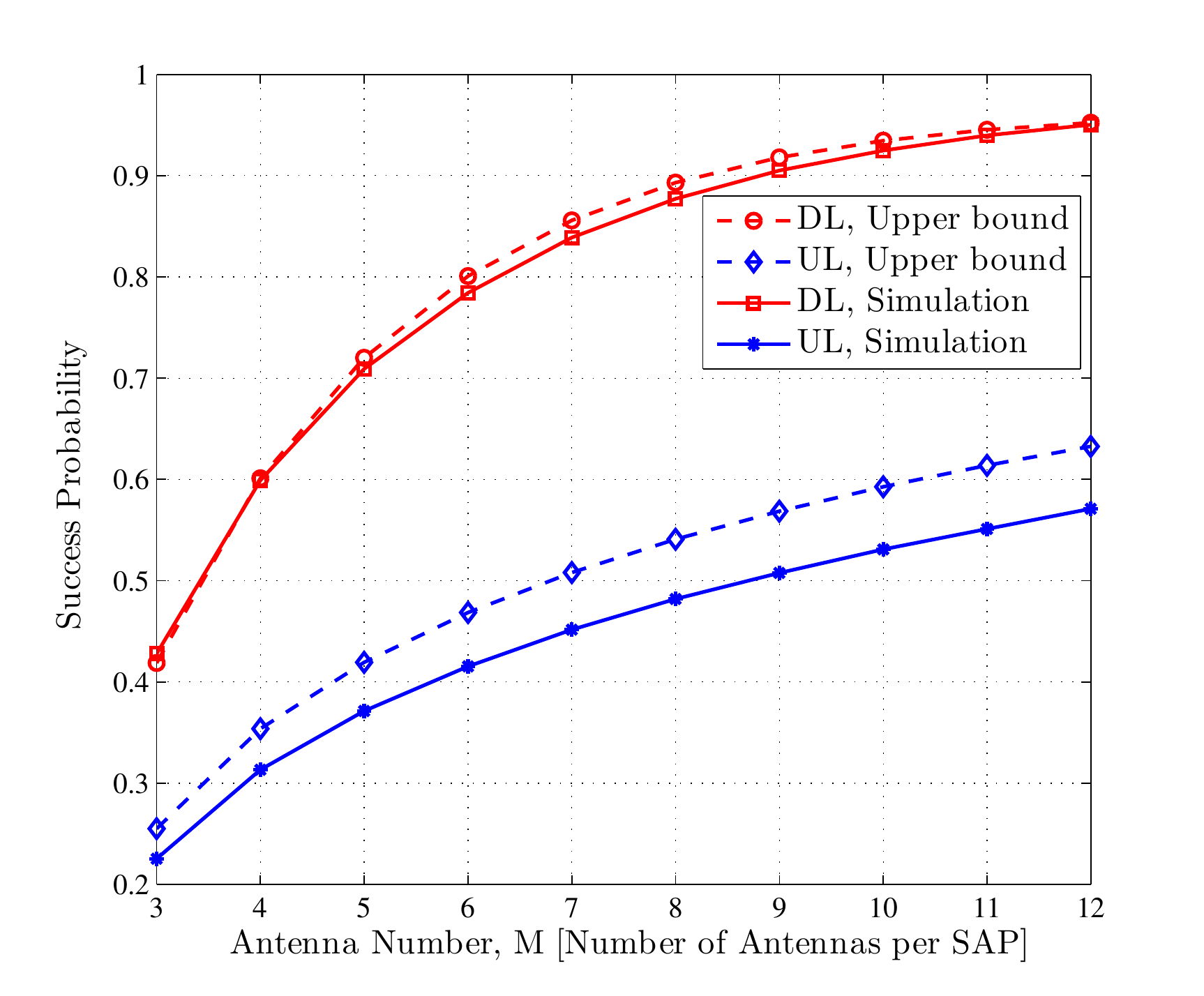}

}

\centering%
\subfloat[]%
{\centering\includegraphics[scale=0.5]{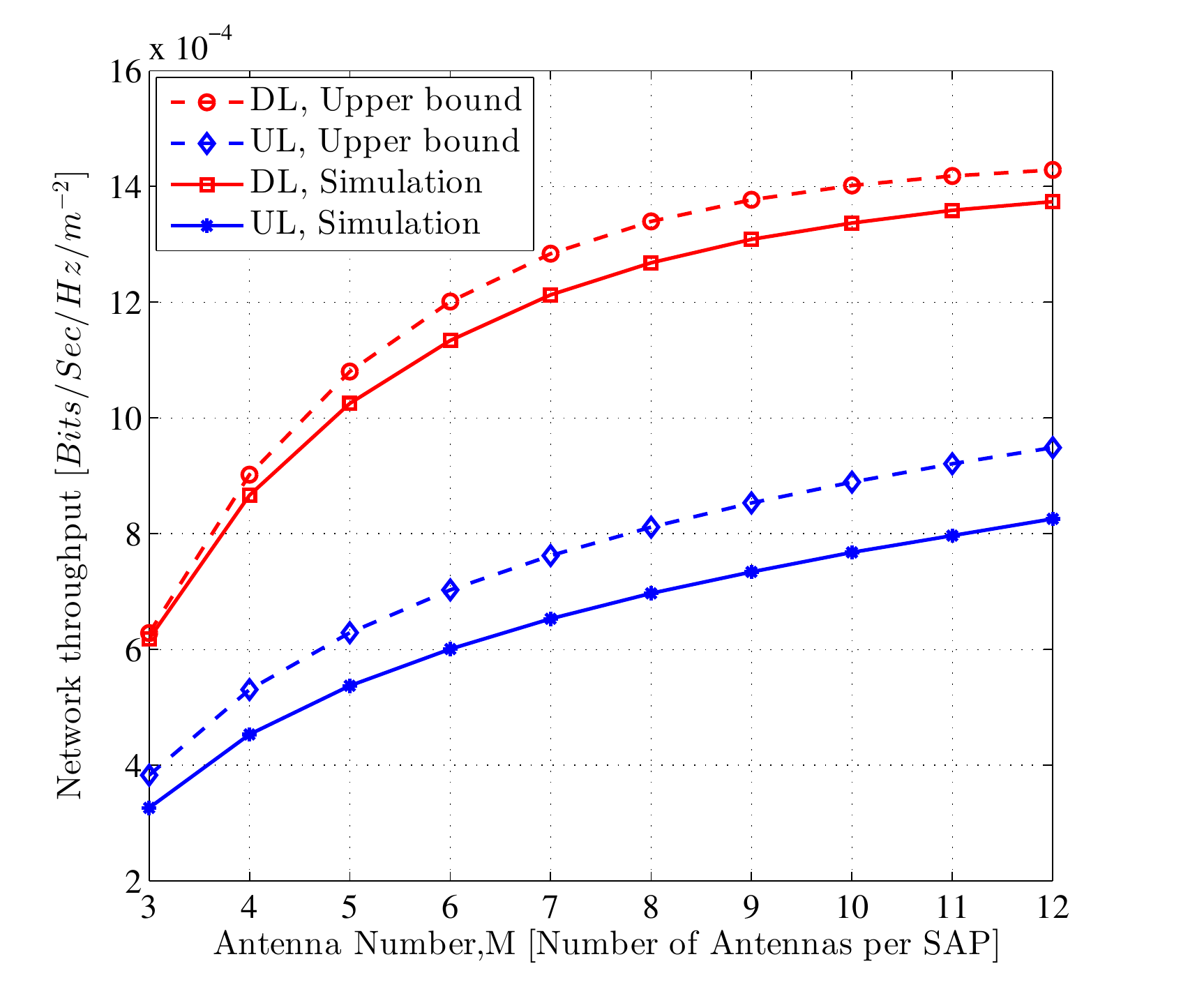}}

\caption{Success probability (a) and network throughput (b) as a function of
SAP antenna number $M$ with DL fraction $p_{\textrm{D}}=0.5$, cluster
size $l=3$, and SINR threshold $\gamma_{\textrm{D}}=\gamma_{\textrm{U}}=0\textrm{dB}$.}

\end{figure}

Figure 3 depicts the success probability and network throughput as
a function of SAP antenna number $M$ where the cluster size is set
to be $l=3$. From Fig. 3(a) and (b), we observe that an increasing
antenna number of SAP $M$ leads to an improvement in both success
probability and network throughput. This can be explained by the enhanced
spatial diversity gain achieved by each MU. With the number of served
MUs $N$ unchanged, a larger antenna number results in the higher
transmit diversity.

\begin{figure}[t]
\centering\includegraphics[scale=0.5]{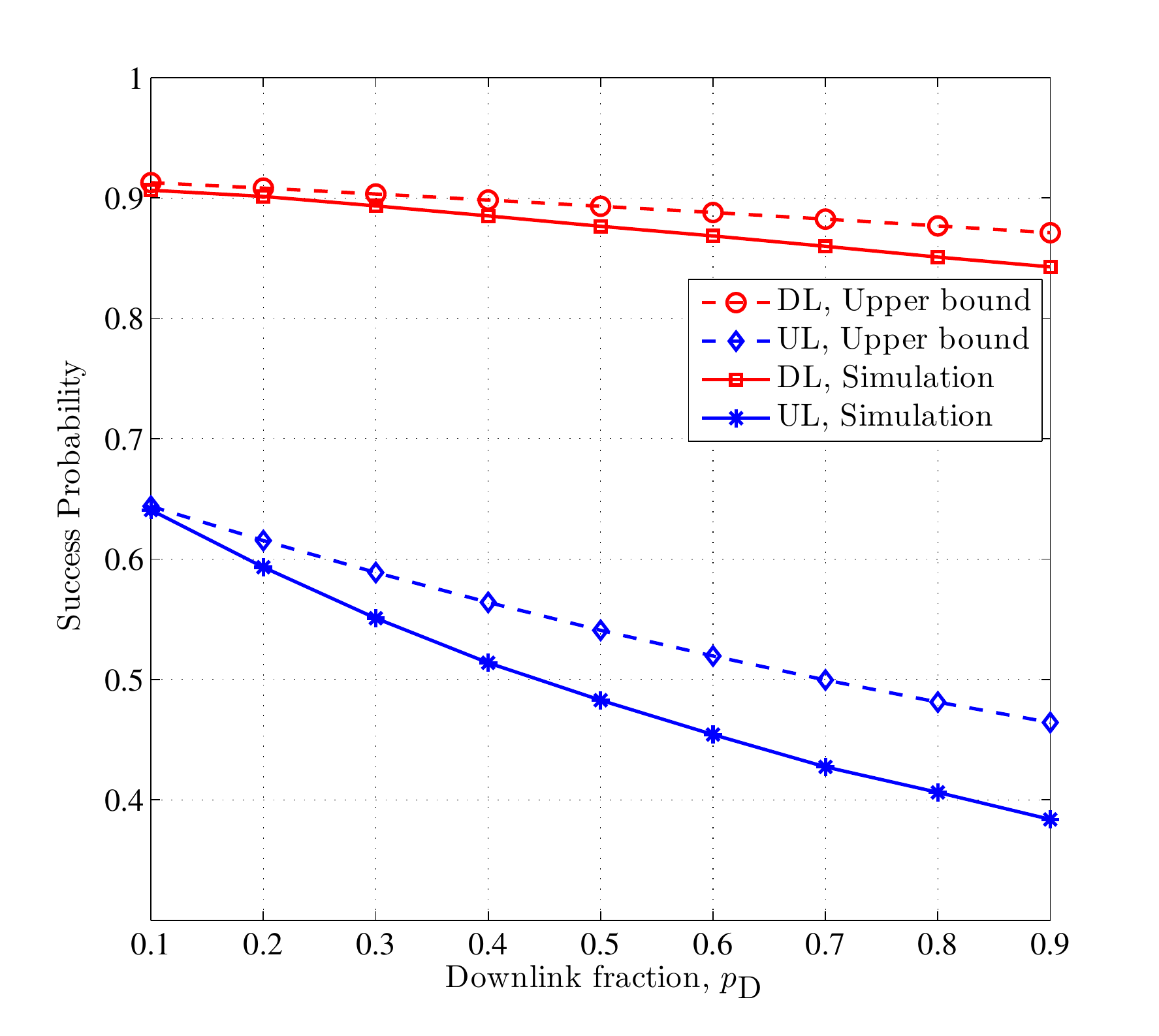}

\caption{Success probability as a function of DL fraction $p_{\textrm{D}}$
with cluster size $l=3$, antenna number per SAP $M=8$, and SINR
threshold $\gamma_{\textrm{D}}=\gamma_{\textrm{U}}=0\textrm{dB}$.}
\end{figure}

In Fig. 4, we plot the success probability as a function of DL fraction
$p_{\textrm{D}}$. As $p_{\textrm{D}}$ grows, the DL interfering
SAP density increases while the UL interfering MU density decreases.
With the considered parameter settings, it shows that both DL and
UL success probabilities decrease with $p_{\textrm{D}}$. This can
be explained by the huge difference in transmit power between SAP
and MU, which results in the incremental aggregated interference.

\begin{figure}[t]
\centering%
\subfloat[]%
{\centering\includegraphics[scale=0.48]{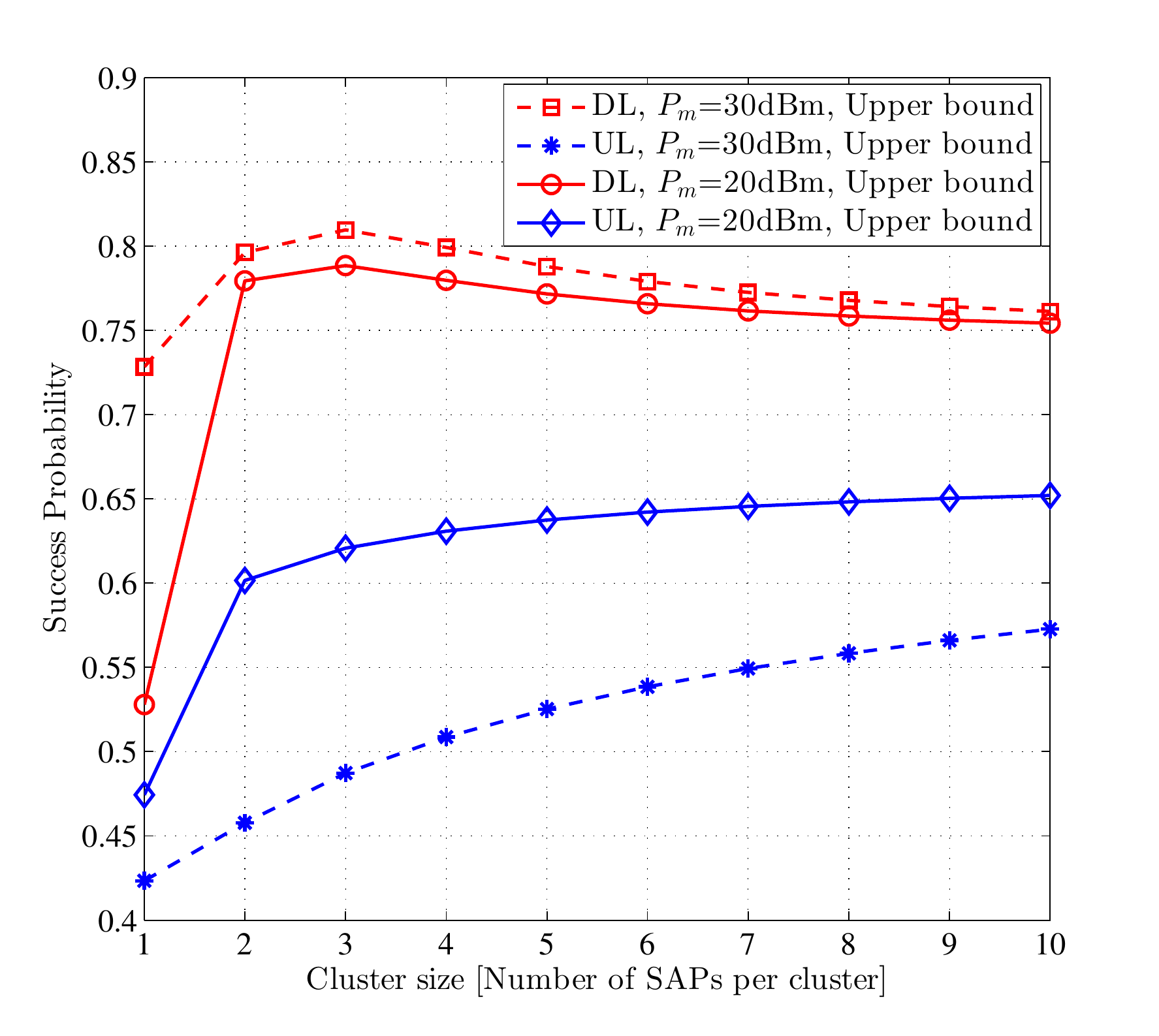}

}

\centering%
\subfloat[]%
{\centering\includegraphics[scale=0.48]{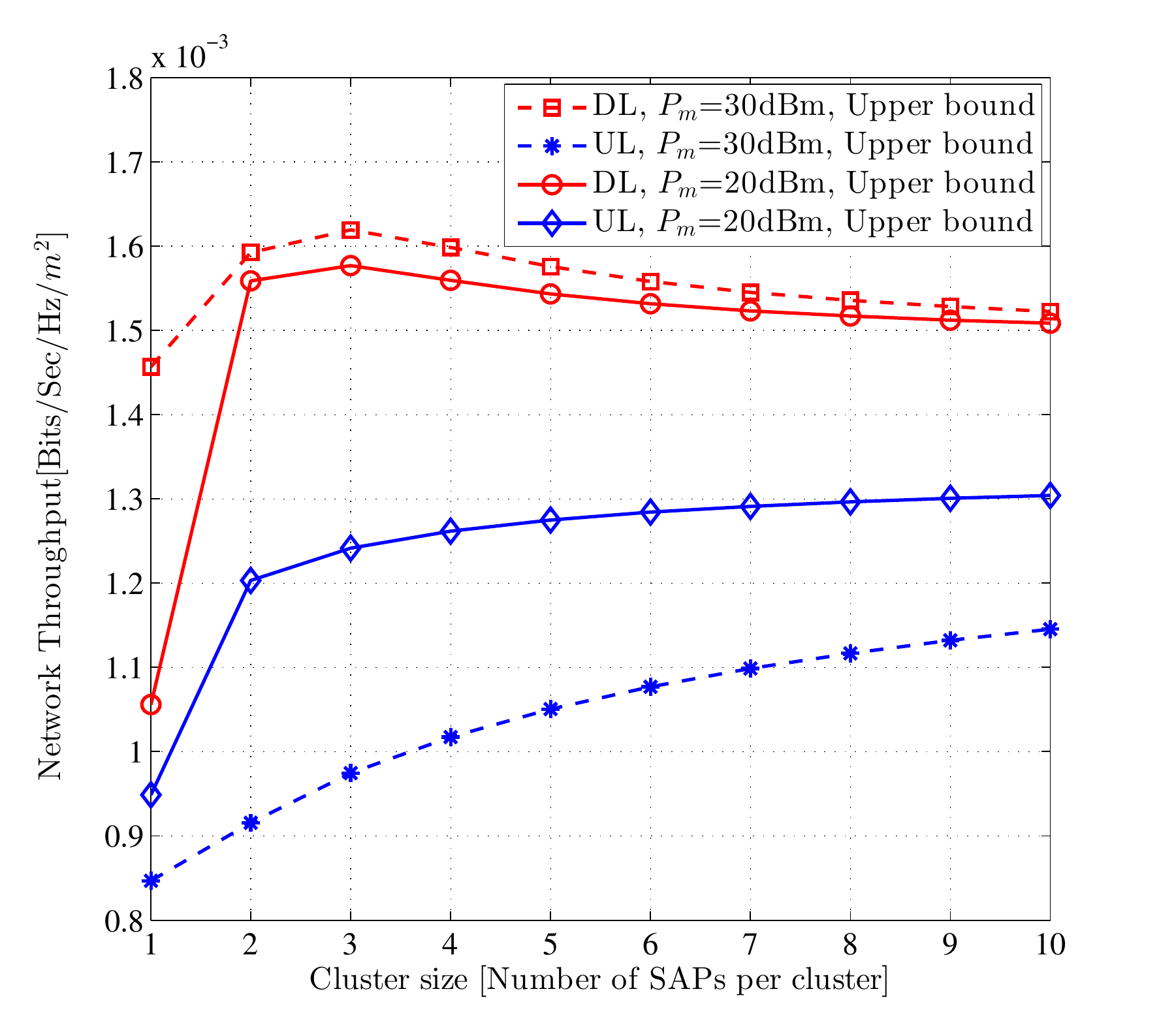}

}

\caption{Success probability (a) and network throughput (b) as a function of
cluster size with antenna number per SAP $M=8$, maximum number of
MUs associated to each SAP $N=4$, DL fraction $p_{\textrm{D}}=0.5$,
and SINR threshold $\gamma_{\textrm{D}}=\gamma_{\textrm{U}}=0\textrm{dB}$.}
\end{figure}

Figure 5 depicts the success probability and network throughput as
a function of cluster size $l$, i.e., the average number of SAPs
per cluster, under different SAP transmit power. We observe that both
success probability and network throughput in DL direction first grows
and then decreases with respect to cluster size. This can be explained
by the fact that when the cluster is small, the inter-cluster MU-to-MU
interference dominates the DL transmission, and an enlargement of
cluster eliminates the MU-to-MU interference, thus, boosting up the
DL performance. However, as the cluster size keeps growing, the intra-cluster
SAP-to-MU interference begins to dominate the DL transmission. As
more and more SAPs are clustered and configured into the same direction,
the increasing intra-cluster DL interference degrades the DL performance.
Fig. 5 also shows that the UL performance always benefits from the
growing cluster. This comes from the fact that the UL performance
is dominated by the severe SAP-to-SAP interference, and an increase
in cluster size helps eliminate the SAP-to-SAP interference, which
improves the UL performance. What's more, a lower $P_{m}$ leads to
an enhancement in UL performance and a decrease in DL performance.
Since $l=1$ corresponds to the traditional D-TDD scheme, Fig. 5(a)
and (b) show the superiority of clustered D-TDD scheme over the traditional
D-TDD scheme in both success probability and network throughput. It
further reveals that our proposed analytical framework can be used
to determine the optimal cluster size (e.g., $l=3$ in this example)
for the DL performance.

\section{Conclusions}

In this work, we proposed an analytical framework to investigate the
network performance of multi-antenna D-TDD small cell network with
cell clustering as the interference mitigation scheme. The framework
allows to evaluate the success probability and network throughput
by accounting for the UL/DL configuration, cluster size, antenna number,
transmit power, SINR threshold, and maximum number of MUs associated
to each SAP. The superiority of the clustered D-TDD over the traditional
D-TDD is revealed by numerical results. The proposed analytical framework
can be used to find the optimal cluster size for the DL network performance,
and it also revealed that a larger cluster size leads to better UL
network performance.

\appendix{}

\emph{A. Proof of Theorem 1}

The success probability $\mu_{\textrm{TX}}$, $\textrm{TX}\in\{\textrm{D,U}\}$
is affected by not only the interference, cluster size, but also the
antenna number. For an SAP associating with $n_{0}$ MUs, the DL success
probability can be derived as{\small{}
\begin{eqnarray}
\mu_{\textrm{D}}(n_{0}) & \overset{(a)}{=} & \int_{0}^{\rho}\textrm{Pr}\bigl(h_{0,0}^{\mathtt{\textrm{D}}}>sI_{IN}^{\textrm{D}}|n_{0}\bigr)f_{r}(r_{0})\mathrm{d}r_{0}\nonumber \\
 & \overset{(b)}{=} & \int_{0}^{\rho}\sum_{i=0}^{M-n_{0}}\frac{1}{i!}\mathbb{E}_{\{I_{IN}^{\textrm{D}}\}}\bigl[(sI_{IN}^{\textrm{D}})^{i}e^{-sI_{IN}^{\textrm{D}}}\bigr]f_{r}(r_{0})\mathrm{d}r_{0}\nonumber \\
 & \overset{(c)}{=} & \int_{0}^{\rho}\sum_{i=0}^{M-n_{0}}\frac{1}{i!}(-s)^{i}\frac{d^{i}}{ds^{i}}\mathcal{L}_{I_{IN}^{\textrm{D}}}(s)\cdot f_{r}(r_{0})\mathrm{d}r_{0},\label{eq:Service_rate_DL}
\end{eqnarray}
}where $\rho$ is cluster radius, $f_{r}(r_{0})=2\pi\lambda_{\textrm{s}}r_{0}\exp(-\pi\lambda_{\textrm{s}}r_{0}^{2})$
is the PDF of the typical link length, and $I_{IN}^{\textrm{D}}\triangleq I_{\mathtt{\textrm{D}}}+\sigma^{2}$.
Step (a) is derived by defining $s=\frac{\gamma_{\mathtt{\textrm{D}}}r_{0}^{\alpha}}{P_{\mathrm{s}}}$,
(b) follows from the CCDF of a Gamma variable $X\sim\Gamma(k,\theta)$,
and (c) is derived by substituting $\mathbb{E}_{X}[X^{n}e^{-sX}]=(-1)^{n}\frac{\mathrm{d}^{n}}{\mathrm{d}s^{n}}\mathcal{L}_{X}(s)$
and $h_{0,0}^{\mathtt{\textrm{D}}}\sim\Gamma(M-n_{0}+1,1)$.

To derive $\mu_{\textrm{D}}(n_{0})$, we need to first calculate the
$i$-th derivative of the Laplace transform of $I_{IN}^{\textrm{D}}$.
Due to the independence of $I_{\mathtt{\textrm{D}}}$ and $\sigma^{2}$,
we have $\mathcal{L}_{I_{IN}^{\textrm{D}}}(s)=\mathbb{E}\Bigl[e^{-s\sigma^{2}}\Bigr]\mathcal{L}_{I_{\mathtt{\textrm{D}}}}(s).$
Conditioned on the typical link length $r_{0}$, the Laplace transform
of $I_{\textrm{D}}$ can be derived as{\small{}
\begin{align}
\mathcal{L}_{I_{\textrm{D}}}(s) & =\mathbb{E}\biggl[\exp\Bigl(-s\sum_{n=1}^{N}\sum_{x_{i}\in\Phi_{\textrm{s}_{n}}^{\textrm{in}}\backslash\{0\}}P_{\mathrm{s}}\parallel x_{0,i}\parallel^{-\alpha}g_{\mathbf{x}_{i},\mathrm{SAP}}^{\textrm{D}}\Bigr)\nonumber \\
 & \times\exp\Bigl[-s\sum_{k=1}^{\infty}\Bigl(\sum_{n=1}^{N}\sum_{x_{j}\in\Phi_{\textrm{s}_{n}}^{\textrm{out}_{k}}}\boldsymbol{1}_{\{\textrm{TX=D}\}}^{\textrm{out}_{k}}P_{\mathrm{s}}\parallel x_{0,j}\parallel^{-\alpha}g_{\mathbf{x}_{j},\mathrm{SAP}}^{\textrm{D}}\nonumber \\
 & +\sum_{n=1}^{N}\sum_{z_{l}\in\Phi_{\textrm{u}_{n}}^{\textrm{out}_{k}}}\boldsymbol{1}_{\{\textrm{TX=U}\}}^{\textrm{out}_{k}}Q_{\mathrm{u}}\parallel z_{0,l}\parallel^{-\alpha}g_{\mathrm{MU}}^{\textrm{D}}\Bigr)\Bigr]\biggr]\nonumber \\
 & =\mathcal{L}_{I_{\textrm{D}}^{\textrm{in}}}(s)\cdot\mathcal{L}_{I_{\textrm{D}}^{\textrm{out}}}(s).
\end{align}
}Specifically, the Laplace transform of the intra-cluster interference
$I_{\textrm{D}}^{\textrm{in}}$ is given by{\small{}
\begin{align}
 & \mathcal{L}_{I_{\textrm{D}}^{\textrm{in}}}(s)\nonumber \\
\stackrel{(a)}{=} & \exp\biggl\{-\sum_{n=1}^{N}\lambda_{s,n}\int_{\mathcal{C}(T,\rho)\backslash\mathcal{B}(0,x_{0})}\bigl(1-\mathcal{L}_{g_{\mathbf{x}_{i},\mathrm{SAP}}^{\mathtt{\textrm{D}}}}(sP_{\mathrm{s}}x^{-\alpha})\bigr)\mathrm{d}x\biggr\}\nonumber \\
\stackrel{(b)}{=} & \exp\biggl\{-2\pi\sum_{n=1}^{N}\lambda_{s,n}\int_{r_{0}}^{\rho}\bigl(1-\frac{1}{(1+sP_{\mathrm{s}}r^{-\alpha})^{n}}\bigr)r\mathrm{d}r\biggr\}\nonumber \\
\stackrel{(c)}{=} & \exp\biggl\{-2\pi\sum_{n=1}^{N}\lambda_{s,n}\sum_{l=1}^{n}C_{n}^{l}\int_{r_{0}}^{\rho}\frac{\bigl(sP_{\mathrm{s}}r^{-\alpha}\bigr)^{l}r}{(1+sP_{\mathrm{s}}r^{-\alpha})^{n}}\mathrm{d}r\biggr\},\label{eq:Lap_IDD_in}
\end{align}
}where $g_{\mathbf{x}_{i},\mathrm{SAP}}^{\textrm{D}}\triangleq\mid\mathbf{h}_{0,i}^{\dagger}\mathbf{W}_{i}\mid^{2}$,
and $\lambda_{\textrm{s},n}=\lambda_{\textrm{s}}f(n)$. Step (a) follows
from the probability generating functional (PGFL) of PPP, (b) follows
from the fact that $g_{\mathbf{x}_{i},\mathrm{SAP}}^{\mathtt{\textrm{D}}}\sim\Gamma(n,1)$,
and (c) comes from Binomial theorem and the change from Cartesian
to polar coordinates.

Similarly, the Laplace transform of the cross-cluster interference
$I_{\textrm{D}}^{\textrm{out}}$ is given by{\small{}
\begin{align}
 & \mathcal{L}_{I_{\textrm{D}}^{\textrm{out}}}(s)\nonumber \\
\stackrel{(a)}{\approx} & \prod_{k=1}^{\infty}\biggl\{ p_{\textrm{D}}\exp\biggl(-2\pi\sum_{n=1}^{N}\lambda_{\textrm{s},n}\sum_{l=1}^{n}C_{n}^{l}\int_{\sqrt{k}\rho}^{\sqrt{k+1}\rho}\frac{(sP_{\mathrm{s}}r^{-\alpha})^{l}r}{(1+sP_{\mathrm{s}}r^{-\alpha})^{n}}\mathrm{d}r\biggr)\nonumber \\
 & +\bigl(1-p_{\textrm{D}}\bigr)\exp\Bigl(-2\pi\sum_{n=1}^{N}\lambda_{\textrm{u},n}\biggl(\Theta\Bigl(\alpha,1,\bigl(sQ_{\textrm{u}}\bigr)^{-1},\sqrt{k+1}\rho\Bigr)\nonumber \\
 & \qquad\qquad\qquad\qquad\qquad-\Theta\Bigl(\alpha,1,\bigl(sQ_{\textrm{u}}\bigr)^{-1},\sqrt{k}\rho\Bigr)\biggr)\biggr\},\label{eq:Lap_IDD_out}
\end{align}
}where $\lambda_{\textrm{u},n}\approx\lambda_{\textrm{s}}nf(n)$,
and {\small{}
\begin{align*}
\Theta\Bigl(\alpha,\beta,u,d\Bigr) & \triangleq\int_{0}^{d}\frac{r^{\beta}}{1+ur^{\alpha}}dr\\
 & =\frac{d^{\beta+1}}{\beta+1}_{2}F_{1}(1,\frac{\beta+1}{\alpha};1+\frac{\beta+1}{\alpha};-ud^{\alpha}),
\end{align*}
}with $_{2}F_{1}(\cdot,\cdot;\cdot;\cdot)$ being the hypergeometric
function. Note that the calculation of $\lambda_{\textrm{u},n}$ follows
from the fact that there is a mapping between an SAP and its associated
MUs. Assume an UL SAP associates with $n$ MUs, then the density of
interfering MUs is equivalent to $\lambda_{\textrm{s}}nf(n)$. Step
(a) is due to the fact that all SAPs in each cluster are simultaneously
configured in DL (resp. UL) with probability $p_{\textrm{D}}$ (resp.
1-$p_{\textrm{D}}$), and the approximation of dividing the cross-cluster
zone into a sequence of disjoint hexagonal rings.

Combining \eqref{eq:Lap_IDD_in} with \eqref{eq:Lap_IDD_out}, we
derive $\mathcal{L}_{I_{IN}^{\textrm{D}}}(s)$ in \eqref{eq:Laplace_DL}{\small{}.
}By substituting \eqref{eq:Laplace_DL} into \eqref{eq:Service_rate_DL},
$\mu_{\textrm{D}}(n_{0})$ is given by \eqref{eq:mu_D}. Following
the similar method, we derive $\mu_{\textrm{U}}(n_{0})$ in \eqref{eq:mu_U}.
This concludes the proof.

\emph{B. Proof of Corollary 1}
\noindent \begin{flushleft}
Define $\gamma(m,z)=\int_{0}^{z}t^{m-1}e^{-t}dt$ as the lower incomplete
Gamma function. The CCDF of Gamma distribution can be expressed as
$\bar{F}_{Z}(z)=1-\frac{\gamma(m,z)}{\Gamma(m)}$. With Alzer's inequality
in \cite{OSIF}, we have 
\begin{equation}
\frac{\gamma(m,z)}{\Gamma(m)}>(1-\exp(-cz))^{m}.
\end{equation}
For $m>1$, we have $c=(\Gamma(m+1))^{-\frac{1}{m}}$. By expanding
the expectation term, we have
\begin{align}
\frac{\mathbb{E}_{I}[\gamma(m,zI)]}{\Gamma(m)} & \geq\mathbb{E}_{I}[(1-\exp(-czI))^{m}]\nonumber \\
 & =\sum_{k=0}^{m}(-1)^{k}C_{m}^{k}\mathcal{L}_{I}(ckz).\label{eq:Alzer_Inequality}
\end{align}
Take $\mu_{\textrm{D}}(n_{0})$ in \eqref{eq:mu_D} as an example,
we have
\begin{align*}
 & \sum_{i=0}^{M-n_{0}}\frac{1}{i!}(-s)^{i}\frac{d^{i}}{ds^{i}}\mathcal{L}_{I_{IN}^{\textrm{D}}}(s)\\
 & =1-\frac{\mathbb{E}_{I_{IN}^{\textrm{D}}}[\gamma(M-n_{0}+1,sI_{IN}^{\textrm{D}})]}{\Gamma(M-n_{0}+1)}
\end{align*}
\begin{align}
 & \leq1-\sum_{i=0}^{\Delta_{n_{0}}}(-1)^{i}C_{\Delta_{n_{0}}}^{i}\mathcal{L}_{I_{IN}^{\textrm{D}}}(i\cdot(\Gamma(\Delta_{n_{0}}+1))^{-\frac{1}{\Delta_{n_{0}}}}s)\nonumber \\
 & \overset{(a)}{=}\sum_{i=1}^{\Delta_{n_{0}}}(-1)^{i+1}C_{\Delta_{n_{0}}}^{i}\mathcal{L}_{I_{IN}^{\textrm{D}}}(i\cdot(\Gamma(\Delta_{n_{0}}+1))^{-\frac{1}{\Delta_{n_{0}}}}s),
\end{align}
where we define $\Delta_{n_{0}}=M-n_{0}+1$ (a) follows from the fact
that $-(-1)^{k}=(-1)^{k+1}$ and that $C_{M-n_{0}+1}^{0}\mathcal{L}_{I_{IN}^{\textrm{D}}}(0)=1$.
This concludes the proof.
\par\end{flushleft}

\bibliographystyle{IEEEtran}
\bibliography{7D__Research_Document_Ongoing_Research_DTDD_New___ng_Draft_Writing_ArXiv_Version_REF_MIMO-TDD}

\end{document}